\documentstyle[12pt]{article}
\begin{document}
\title{A note on the paper, "The Universe of Fluctuations"}
\author{B.G. Sidharth\\
Centre for Applicable Mathematics \& Computer Sciences\\
B.M. Birla Science Centre, Adarsh Nagar, Hyderabad - 500 063 (India)}
\date{}
\maketitle
\footnotetext{Email:birlasc@hd1.vsnl.net.in}
\begin{abstract}
We examine how the consequences which follow from a recent model, both in
cosmology and at the elementary particle level have since been observationally
and experimentally confirmed. Some of the considerations of the model are also
justified from alternative viewpoints. It is also shown how the standard
Big Bang and quark models can be recovered from the above theory.
\end{abstract}
\section{Introduction}
In the above paper hereinafter referred
 to as\cite{r1} (cf. also
references\cite{r2,r3}), it was pointed out that Fermions could be treated
as Kerr-Newman type Black Holes, but with an important Quantum Mechanical
proviso: One has to take into consideration the fact that arbitrarily small space time
intervals are  purely classical concepts, or as pointed out in\cite{r4}, classical
approximations. Strictly speaking there is a space-time quantization
as pointed out in\cite{r1}. That is, there are minimum space time intervals,
viz., the Compton wavelength and time, reminiscent of the earlier concept
of the chronon\cite{r5,r6}. (Moreover, as pointed out in\cite{r7}, this may well
be more fundamental than energy quanta which latter can be shown to follow
from the former.)\\
In this light, the naked singularity which arises in the description of
elementary particles as Kerr-Newman Blak Holes disappears and as shown in
\cite{r1,r2,r3} one gets a rationale for the electron's anomalous gyro magnetic
ratio $g = 2$, the neutrino's handedness and several other otherwise adhoc features
including a possible scheme for the unification of electromagnetism and gravitation.\\
In particular as was seen in\cite{r1} one could get a cosmological scheme
from the above which deduces from theory the values of cosmological quantities
like the age, radius and mass of the universe, the Hubble Constant the
cosmological constant and so on,
as also the well known inexplicable Dirac Large Number coincidences, but also
the equally mysterious Weinberg relation between the pion mass and Hubble
constant (cf. also\cite{r8}), given $N$, the number of baryons as the
only large scale parameter, retaining the well known microphysical constants\cite{r9}.
The model also predicts an ever expanding, accelerating universe.\\
In this note we on the one hand draw attention to some of these and other
consequences of the above model and point out that these have since been
confirmed by observation and experiment. On the other hand, we will show how
standard theory-- the Big Bang model and the quark model-- can be recovered from
the new approach. In the process, we also get a rationale for quantum
non-locality, which no longer shows up as being spooky.
\section{Fluctuational Cosmology}
We briefly recapitulate the main results of the cosmological model described
in reference\cite{r1,r8}. The main concepts are:\\
From a background Zero Point Field, particles, typically pions, are fluctuationally
created within space time intervals ($l,\tau$), the Compton wavelength and
Compton time of the particle. Further given $N$ particles (typically pions) in the universe
at any epoch, $\sqrt{N}$ particles  are fluctuationally created.\\
From here it was shown that the mass and the radius of the universe would
follow via the equations
$$M = Nm,$$
where $m$ is the pion mass, and
$$R = \frac{GM}{c^2}$$
where $M$ and $R$ are the mass and radius of the universe and $N$, the sole
cosmological or large scale parameter is the number of particles $\sim 10^{80}$
in the universe.\\
Next using the fact that
\begin{equation}
\frac{dN}{dt} = \frac{\sqrt{N}}{\tau} = \frac{mc^2}{\hbar} \sqrt{N}\label{e1}
\end{equation}
we deduce on integration from $t = 0, N = 0$ to the present epoch, that,
$$\sqrt{N} = \frac{2mc^2}{\hbar} T$$
where $T$ is the age of the universe and from the above
$\approx 10^{17}secs$., which is correct.\\
From the above it was shown  in\cite{r1} that the Hubble Constant is
given by
$$H = \frac{Gm^3c}{\hbar^2}$$
which is correct, or equivalently
\begin{equation}
m = (\frac{\hbar^2 H}{Gc})^{1/3}\label{e2}
\end{equation}
This has been considered to be a mysterious relation, as pointed out
\cite{r10}, an inexplicable coincidence linking a typical elementary
particle mass to cosmological parameters. However in our model it follows as
a natural consequence to the theory.\\
Moreover it was also shown that we get the cosmological constant also
consistently as
$$\wedge \sim H^2$$
According to this model the universe would accelerate and continue to expand
for ever with ever decreasing density. It is remarkable that these conclusions
have very recently been confirmed by several independent teams of
observers\cite{r11,r12,r13}.\\
All this apart, in the above model the supposedly mysterious large number
coincidences which lead to Dirac's large number hypothesis are actually
consequences of the theory (cf.ref.\cite{r8} also).\\
To get further insight, we observe that, as noted in the introduction, the
concept of space time points is but a classical approximation which has been
critically examined (cf. for example reference\cite{r14}). We argued in
\cite{r1} that as with the earlier concept of a minimum time unit, the
chronon, the Compton time is physically the meaningful concept, as also
the minimum length unit viz., the Compton wavelength.\\
It was suggested that an elementary particle was clearly a fudge, a Kerr-Newman
type Black Hole within the above Compton wavelength, within which
we encounter as is well known Zitterbewegung and superluminal non-local
effects. It was pointed out that the fact that a Dirac particle has luminal
velocity, if treated as a point is symptomatic of the fact that these
Fermions are manifestations of the background Zero Point Field trapped
within the Compton wavelength. Equivalently, it was argued that one could
think of the universe as containing $N$ such luminal instantaneous particles
or Ganeshas which would then have a statistical uncertainity in position
of $\frac{V}{N}, V$ being the total volume of the universe. Whence it was shown
that the typical volume of uncertainity
$$\frac{V}{N} \sim \lambda^3_{\mbox{thermal}} \quad \approx
(\frac{\hbar}{\sqrt{m^2 c^2}})^3
= (\frac{\hbar}{mc})^3,$$
in this case, owing to the luminal (r.m.s) velocity.\\
Indeed, another way of looking at this is, as is known,\cite{r15}, for an
assembly of $N$ particles in a volume of radius $R$, a typical uncertainity
length $l$ is given by
\begin{equation}
l \approx \frac{R}{\sqrt{N}},\label{e3}
\end{equation}
which is indeed true for the pion Compton wavelength. Alternatively, the above
relation can be deduced by using the fact that when the number of
fluctuations in the number of particles is $\sqrt{N}$, the excess
electrostatic potential energy of the electrons, for example is given by
$\frac{e^2\sqrt{N}}{R}$
which is balanced by the electron energy $m_ec^2$ (cf.\cite{r1} for details).\\
It was pointed out that this was a holistic picture: The Quantum Mechanical
uncertainity and therefore the minimum space time intervals are a result
of the universe as a whole.\\
It is possible to come to this conclusion by
yet another route. As is known\cite{r16}, the fluctuation in the mass of a
typical elementary particle like the pion, due to the fluctuation of the
particle number $\sim \sqrt{N}$ is given by
$$\frac{G\sqrt{N} m^2}{c^2R}$$
whence we get
\begin{equation}
(\Delta mc^2) T = \frac{G\sqrt{N}m^2}{R} T = \frac{G\sqrt{N}m^2}{c}\label{e4}
\end{equation}
One can now easily verify that,
\begin{equation}
\hbar \approx \frac{G\sqrt{N}m^2}{c},\label{e5}
\end{equation}
so that (\ref{e4}) gives us the well known Quantum Mechanical equation
$$\Delta E \Delta t \approx \hbar.$$
Equation (\ref{e5}) gives us the Quantum Mechanical Planck constant $\hbar$ in terms of
of classical quantitites, more accurately, in terms of the classical
fluctuation in particle number $\sqrt{N}$. Further (\ref{e5}), as can be verified
is equivalently, the well known electgromagnetism/gravitational interaction
ratio
\begin{equation}
e^2/Gm^2 \sim 10^{-40}\label{e6}
\end{equation}
on using the relation $c\hbar = 137e^2$ which again is nothing but the equation (\ref{e3}).\\
In other words the Heisenberg uncertainity relation for energy and time
period follows from the fluctuation in particle number. Similarly the
corresponding Heisenberg relation for momentum and distance can also be
shown to hold.\\
In this light of what has been called the micro-macro nexus in\cite{r1} the
mysterious Weinberg relation (\ref{e2}) is perfectly meaningful. Similarly also
the curious fact that, as pointed out in\cite{r17} the radiation time for
a pion equals the age of the universe. Infact it can be seen why the large
number coincidences including (\ref{e6}), are no coincidences at all: All
these express the fact that the small is tied up with the large by the mechanism
discussed above.\\
It has also been argued by the author (cf.ref.\cite{r7}) that in this light
the supposedly spooky quantum non-locality, or the breakdown of local
realism is physically meaningful. The point is, and we emphasize this, the
space time measurements we make with the minimum uncertainity intervals, are the physically
meaningful measurements, and these are holistic: Local realism is a purely
classical concept and as such is approximate. Indeed if in (\ref{e3}) $N \to \infty$,
then $l \to 0$ and we have the classical space-time points.\\
Once it is recognized that our classical concept of space time is an approximation with the
minimum space time interval $\to 0$, it is easy to
reconcile our theory with a big bang scenario. Infact in equation (\ref{e1}),
if $\tau \to 0$ we get a singular creation-- all particles being created
momentarily as $\frac{dN}{dt} \to \infty$.\\
To examine this situation in greater detail we observe that, as noted by
several authors\cite{r18,r19}, the minimum space time intervals are at the
Planck scale. Indeed, at the Planck scale Quantum Mechanics and classical
General Relativity meet\cite{r20}. An easy way in which this can be seen is
by considering a Planck mass $m_P \sim 10^{-5}gms$ for which we have
\begin{equation}
\frac{Gm_P}{c^2} = \hbar /m_Pc\label{e7}
\end{equation}
The left side of (\ref{e7}) represents the classical Schwarzchild Black Hole radius while
the right side is the Quantum Mechanical Compton wavelength (ofcourse, these
particles are too massive to be termed really elementary, and moreover they
are too short lived with life times $\sim 10^{-42}secs$, that is the Planck
time).\\
Another way of looking at (\ref{e7}) is that at the Planck scale, it can also be
written as
$$\frac{Gm^2_P}{e^2} \approx 1,$$
This shows that the entire energy is gravitational as compared to the usual
equation (\ref{e6}).\\
We now use the fact that our minimum space time intervals are $(l_P, \tau_P)$,
the Planck scale, instead of $(l, \tau)$ of the pion, as above (cf.also Section
4).\\
With this new limit, it can be easily verified that the total mass in the volume
$\sim l^3$ is given by
\begin{equation}
\rho_P \times l^3 = M\label{e8}
\end{equation}
where $\rho_P$ is the Planck density and $M$ as before is the mass of the
universe.\\
Moreover the number of Planck masses in the above volume $\sim l^3$ can
easily be seen to be $N' \sim 10^{60}$. However, it must be remembered that
in the physical time period $\tau$, there are $10^{20}$ (that is $\frac{\tau}
{\tau_P})$ Planck life times. In other words the number of Planck particles
in the physical interval $(l, \tau)$ is $N \sim 10^{80}$, the total particle
number.\\
That is from the typical physical interval $(l, \tau)$ we recover the entire
mass and also the entire number of particles in the universe, as in the Big
Bang theory. This also provides the explanation for the above puzzling
relations like (\ref{e8}).\\
That is the Big Bang theory is a characterization of the new model in the
classical limit at Planck scales.
\section{The Quark Model}
In reference\cite{r1}, our starting point for the description of an elementary
particle in classical terms was from the linearized theory of General
Relativity viz., (cf. ref.\cite{r4}),
\begin{equation}
g_{\mu v} = \eta_{\mu v} + h_{\mu v}, h_{\mu v} = \int \frac{4T_{\mu v}
(t - |\vec x - \vec x'|,\vec x')}{|\vec x - \vec x'|} d^3 x'\label{e9}
\end{equation}
with the usual notation.\\
As pointed out in references\cite{r1,r2,r3}, starting from (\ref{e9}) and treating
the Compton wavelength as the boundary we can recover the spin half, the
gravitational potential and the electrostatic potential, as also the relation
(\ref{e6}).
Specifically the electrostatic potential is given by
\begin{equation}
\frac{ee'}{r} = A_0 \approx \frac{2\hbar G}{r}\int \eta^{\mu \nu}
\frac{d}{d\tau}T_{\mu \nu} d^3 x'\label{e10}
\end{equation}
where $e'$ is the test charge.\\
In all these cases we consider, distances large compared to the Compton
wavelength. But, as was shown (cf.ref.\cite{r3}), for distances comparable to the Compton wavelength
(\ref{e9}) leads to a QCD type potential viz.,
\begin{eqnarray}
4 \int \frac{T_{\mu \nu} (t,\vec x')}{|\vec x - \vec x' |} d^3 x' +
(\mbox terms \quad independent \quad of \quad \vec x), \nonumber \\
+ 2 \int \frac{d^2}{dt^2} T_{\mu \nu} (t,\vec x')\cdot |\vec x - \vec x' |
d^3 x' + 0 (| \vec x - \vec x' |^2) \propto - \frac{\propto}{r} + \beta r\label{e11}
\end{eqnarray}
This is suggestive of the fact that the Quark model also would follow from the
above considerations. We will show that this is indeed the case. First we will
see how the puzzling $1/3$ and $2/3$ charges of the quarks emerge.\\
Our starting point is the expression for the electrostatic potential(\ref{e10}).\\
We first note that the electron's spin half which is correctly
described in the above model of the Kerr-Newman Black Hole, outside the
Compton wavelength  automatically implies three spatial dimensions\cite{r21,r22}.
This is no longer true as we
approach the Compton wavelength in which case we deal with low space
dimensionality\cite{r23}. This indeed has been already
observed in experiments with nanotubes\cite{r24,r25}. In other words for the Kerr-Newman
Fermions spatially confined to distances of the order of their Compton
wavelength or less, we actually have to consider two and one spatial
dimensionality.\\
Using now the well known fact\cite{r26} that each of the $T_{\imath j}$ in
(\ref{e10}) or (\ref{e11}), is given by $\frac{1}{3} \epsilon, \epsilon$
being the energy density, it follows from (\ref{e10}) that the
particle would have the charge $\frac{2}{3} e$ or $\frac{1}{3}e$, as in the
case of quarks.\\
Moreover, as noted earlier (cf.ref.\cite{r3} also), because we are
at the Compton wavelength scale, we encounter predominantly the "negative
energy" components
$\chi$ of the Dirac four spinor $\left(\begin{array}{c}\chi \\ \psi 
\end{array}\right)$ with opposite parity.
So, as with neutrinos (discussed in ref.\cite{r3}),
this would mean that the quarks would display helicity, which indeed is true:
As is well known, in the $V-A$ theory, the neutrinos and relativistic quarks
are lefthanded while the corresponding anti-particles are right handed (brought
out by the small Cabibo angle)\cite{r27}.\\
All this also automatically implies that these fractionally
charged particles cannot be observed individually because they are by their
very nature spatially confined. This is also expressed by the confining part
of the QCD potential (\ref{e11}). We come to this aspect now.\\
Let us consider the QCD type potential (\ref{e11}). To facilitate comparison
with the standard literature\cite{r28}, we multiply the left hand expression
by $\frac{1}{m}$ (owing to the usual factor $\frac{\hbar^2}{2m})$ and also
go over to natural units $c = \hbar = 1$ momentarily. The potential then
becomes,
\begin{equation}
\frac{4}{m} \int \frac{T_{\mu v}}{r} d^3 x + 2m \int T_{\mu v} r d^3x \equiv
-\frac{\propto}{r} + \beta r\label{e12}
\end{equation}
Owing to the well known relation (cf.ref.\cite{r4})
$$Gm = \int T^{00}d^3x$$
$\propto \sim O(\ref{e1})$ and $\beta \sim O(m^2),$
where $m$ is the mass of the quark. This is indeed the case for the QCD
potential (cf.ref.\cite{r28}). Interestingly, as a check, one can verify
that, as the Compton wavelength distance $r \sim \frac{1}{m}$ (in natural
units), the energy given by (\ref{e12}) $\sim O(m)$, as it should be.\\
Thus both the fractional quark charges (and handedness) and their masses are seen to arise from
this formulation.\\
To proceed further we consider (\ref{e10}) (still remaining in natural units):
\begin{equation}
\frac{e^2}{r} = 2Gm_e \int \eta^{\mu v}\frac{T_{\mu v}}{r}d^3x\label{e13}
\end{equation}
where at scales greater than the electron Compton wavelength, $m_e$ is the
electron mass. At the scale of quarks we have the fractional charge and
$e^2$ goes over to $\frac{e^2}{10} \approx \frac{1}{1370} \sim
10^{-3}$.\\
So we get from (\ref{e13}),
$$\frac{10^{-3}}{r} = 2Gm_e \int \eta^{\mu v} \frac{T_{\mu v}}{r} d^3 x$$
or,
$$\frac{\propto}{r} \sim \frac{1}{r} \approx 2G.10^3 m_e \int \eta^{\mu v}
\frac{T_{\mu v}}{r} d^3x$$
Comparison with the QCD potential (\ref{e12}) shows that the now fractionally charged
Kerr-Newman fermion, viz the quark has a mass $\sim 10^3 m_e$, which is
correct.\\
If the scale is such that we do not go into fractional charges, we
get from (\ref{e13}), instead, the mass of the intermediary particle as
$274 m_e,$ which is the pion mass.\\
All this is ofcourse completely consistent with the physics of strong
interactions.
\section{Discussion}
We have seen above that the Compton wavelength and time on the one hand and
the pion on the other have played a crucial role, and we have argued that
this is not accidental.\\
Indeed, it can be shown that the Compton wavelength and time are the critical
scales at which particle creation takes place, and moreover our earlier
conclusion that at about the Planck mass, the number of particles $\sim 10^{80}$
(the number of elementary particles) can also be deduced from considerations
of $GUT$\cite{r29}. All this throws further light on (\ref{e8}) and the
subsequent considerations.\\
As for the role of the pion as a typical elementary particle, this has been
recognized because of its role in strong interactions. Interestingly, very
much in the spirit of the above considerations of particle creation from a
background ZPF, the critical mass at which a background photon vapour
condenses into particles is of the order of the pion mass itself\cite{r30}.\\
Finally it may be remarked that our conclusion in Section 3 that at the
Compton wavelength scale, and therefore, as argued, in our model, in two and
one dimensions, handedness shows up, can be easily verified from the
relativistically covariant equations in these dimensions\cite{r31}.

\end{document}